\documentstyle[aps,multicol,prl,amsfonts]{revtex}
\tighten
\begin{document}
\draft
\title{Quantum-Classical Complexity-Security Tradeoff In Secure Multi-Party
 Computation}
\author{H. F. Chau\footnote{e-mail: hfchau@hkusua.hku.hk}}
\address{Department of Physics, University of Hong Kong, Pokfulam Road, Hong
 Kong}
\date{\today}
\maketitle
\begin{abstract}
 I construct a secure multi-party scheme to compute a classical function by a
 succinct use of a specially designed fault-tolerant random polynomial quantum
 error correction code. This scheme is secure provided that (asymptotically)
 strictly greater than five-sixths of the players are honest. Moreover, the
 security of this scheme follows directly from the theory of quantum error
 correcting code, and hence is valid without any computational assumption. I
 also discuss the quantum-classical complexity-security tradeoff in secure
 multi-party computation schemes and argue why a full-blown quantum code is
 necessary in my scheme.
\end{abstract}
\pacs{PACS numbers: 03.67.Dd, 03.67.Lx, 03.67.Hk, 89.70.+c}
\begin{multicols}{2}
\section{Introduction}
\label{S:Intro}
 Quantum computers are more powerful than classical computers in a number of
 applications such as integer factorization \cite{Shor_Alg}, database search
 \cite{Grover_Search} and secret key distribution \cite{QCrypto,qkd}. Besides,
 careful use of entanglement reduces the multi-party communication complexity
 of certain functions \cite{Comm_Comp} and allows secret sharing \cite{QSS}. On
 the other hand, certain post-modern cryptographic applications, including bit
 commitment \cite{Bit_Comm} and ideal two-party secure computation
 \cite{2_party} are impossible if the cheater has a quantum computer. Thus, it
 is important to investigate the power and limitation of quantum computers.
 Moreover, the quantum versus classical and security versus complexity
 tradeoffs for certain multi-party computational tasks deserve in-depth study.
\par
 In this Paper, I analyze the quantum versus classical and security versus
 complexity tradeoffs in secure multi-party computation. In secure multi-party
 computation, $n$ players each with a private classical input $x_i$ want to
 compute a commonly agreed classical function $z=f(x_1,x_2,\ldots ,x_n)$ in
 such a way that (i)~all players either know the value of $z$ or abort after
 detecting a cheater/eavesdropper, (ii)~no one can gain information on the
 private input of an honest player except those logically following $z$, and
 (iii)~a limited number of cheating players cannot alter the final outcome $z$.
 Moreover, the above three conditions hold even if all cheaters and
 eavesdroppers cooperate.
\par
 Secure multi-party computation can be used as a basic building block for a
 number of extremely useful protocols including secure election and anonymous
 messages broadcast. Thus, it is important to devise a secure multi-party
 computation scheme that tolerates as many cheaters as possible on the one
 hand, and requires as few communication between the players on the other.
\par
 Several classical secure multi-party computation schemes existed in
 literature. The security of some of these schemes \cite{mental_game} are based
 on either the security of certain (classical) oblivious transfer or
 (classical) bit commitment protocols. Hence, their methods are insecure if a
 cheating player has unlimited computational power. Later on, Ben-Or {\em et
 al.} \cite{nk_func} and Chaum {\em et al.} \cite{1way_func} independently
 proposed multi-party computation methods based on a distributed computing
 version of the so-called $(k,n)$-secret sharing scheme \cite{nk_scheme}. Their
 schemes are unconditionally secure provided that less than one third players
 cheat. This is true even when the cheaters cooperate. Besides, the one third
 cheating player bound is tight among all classical protocols which allow
 secret communications between any two players \cite{nk_func}. Later on, Rabin
 and Ben-Or showed that if each player can broadcast a message to all other
 players and that each pair of players can communicate secretly, then there is
 an unconditionally secure way to compute $z$ if less than a half of the
 players cheat \cite{verify}. The one half cheating player bound is tight
 among all classical schemes which allow secret communications between
 any two players as well as public broadcasting \cite{verify}.
\par
 How much resources is required in classical conditionally secure multi-party
 computation? In all classical schemes known to date, the $n$ players must
 communicate securely with others. Hence, $n(n\!-\!1)/2$ classical secure
 communication channels are required. Suppose each player has a private input
 of length $k$, then initially, they have to distribute their private inputs
 via certain secret sharing schemes. To do so, each player has to send out
 $\mbox{O}(n k)$ bits. Thus, $\mbox{O}(n^2 k)$ bits of (secret) classical
 communications are necessary for the initial setup in the whole system. To
 perform distributed computation, up to $\mbox{O} (n^2 k)$ bits of (secret)
 communications and computation per arithmetical operation are required
 \cite{nk_func,verify}. In addition, to verify that every player's secret input
 is correctly distributed in the secret sharing scheme, an extra $\mbox{O}
 (n^3 k)$ bits of communications are needed \cite{nk_func,1way_func,verify}.
 Since the number of secret communication channels scales quadratically with
 the number of players, classical secure multi-party computation is rarely used
 in practice for more than, say, ten players \cite{Applied_Cry}. In fact, the
 classical schemes by Ben-Or {\em et al.} and Chaum {\em et al.}, being
 generic, are design primarily to point out the plausibility of secure
 multi-party computation.
\section{The Quantum Secure Multi-Party Computation Scheme}
\label{S:Scheme}
 Now, let me report a quantum secure multi-party computation scheme that
 requires fewer communication channels and resources at the expense of
 tolerating fewer cheaters. Without lost of generality, I may assume that the
 private input for each player as well as the output of the function $f$ are
 chosen from a finite field ${\mathbb F}_q$ some prime $q$. My scheme goes as
 follows:
\begin{enumerate}
 \item All players agree on a common computational basis for quantum
  computation, an exponentially small security parameter $e > 0$, as well as
  two random polynomial quantum error correcting codes (QECCs) $C_1$ and $C_2$
  \cite{random}. In particular, they choose $C_1$ to be the $[[n,1,d]]_q$ code
  where the prime $q > n$, and $3d\leq n\!+\!2$. More precisely, $C_1$ encodes
  each $q$ary quantum register $|a_0\rangle$ into $n$ $q$ary quantum registers
  $\sum_{a_1,a_2,\ldots ,a_{d-1} = 0}^{q-1} \bigotimes_{i=1}^n |a_0 + a_1 y_i +
  a_2 y_i^2 + \cdots + a_{d-1} y_i^{d-1}\rangle / q^{(d-1)/2}$ where $y_i$ are
  distinct non-zero elements in ${\mathbb F}_q$. The distance of this code is
  $d$ and hence it can correct up to $\delta \equiv \left\lfloor \frac{d-1}{2}
  \right\rfloor$ errors.\footnote{The distance of this code is less than that
  reported in Ref.~\cite{random}. Nonetheless, I still call this a random
  polynomial code because this code closely resembles that reported in
  Ref.~\cite{random}.} Furthermore, I denote the $[[n,1,d]]_q$ QECC
  $|a_0\rangle \longmapsto \sum_{a_1,a_2,\ldots ,a_{n-d+1}=0}^{q-1}
  \bigotimes_{i=1}^n |a_0 + a_1 y_i + a_2 y_i^2 + \cdots +a_{n-d+1}
  y_i^{n-d+1}\rangle / q^{(n-d+1)/2}$ by $\tilde{C}_1$. In addition, $C_2$ is
  chosen to be the $[[4d'\!+\!1,1,2d'\!+\!1]]_q$ random polynomial QECC
  \cite{random} whose fidelity of quantum computation using imperfect devices
  is greater than $1\!-\!e$. (Since the random polynomial QECC $C_2$ has a
  fault-tolerant implementation \cite{random}, thus, by concatenate coding, the
  threshold theorem in fault-tolerant quantum computation guarantees the
  existence of such a QECC $C_2$ \cite{random,ftqc,Preskill}.) As we shall see
  later on, the choice of the value of the distance $d$ only affect the number
  of cheaters that can be tolerated by the scheme. \label{Alg:Code_Selection}
 \item Each player sets up a quantum channel with a central routing station.
  He/She may establish relay stations along each quantum channel in such a way
  that the noise level in each quantum channel segment is small enough to
  perform entanglement purification. (See
  Refs.~\cite{repeater,purification,n_pure} for details.) Furthermore, each
  player also has access to a classical public unjammable channel for
  broadcasting. \label{Alg:Relay_Selection}
 \item The players, central routing channel and relay stations separately
  prepare a few copies of the state $|\Phi\rangle \equiv \sum_{k=0}^{q-1}
  |kk\rangle / \sqrt{q}$. They encode each copy using QECC $C_2$, and share
  these encoded state $|\Phi\rangle$ between the two ends of each quantum
  communication channel segment. Then, they perform fault-tolerant entanglement
  purification procedure as discussed in Refs.~\cite{purification,n_pure} on
  these shared states. Afterwards, these possibly impure encoded states
  $|\Phi\rangle$ shared between each channel segment from one player to another
  are connected together by quantum teleportation \cite{qkd,repeater,teleport}.
  Finally, each pair of players test the purity of their shared encoded states
  $|\Phi\rangle$ by a variation of the fault-tolerant random hashing technique
  described in Ref.~\cite{qkd}. (Readers may refer to
  Appendices~\ref{S:Teleport_Note} and~\ref{S:Hashing_Note} for detail
  description of the teleportation and the random hashing procedures,
  respectively.) They proceed to step~\ref{Alg:Secret_Dist} only if the random
  hashing test is passed for each pair of players. And in this case, each pair
  of players will share a number of almost perfect encoded logical state $|\Phi
  \rangle$. The entanglement shared between each pair of players in this way
  can then be used to securely transport states among themselves in
  step~\ref{Alg:Secret_Dist}. Clearly, shared $|\Phi\rangle$ is not the only
  possible way to establish such an entanglement. In fact, one may replace the
  state $|\Phi\rangle$ in this scheme by an EPR pair. Nevertheless, the scheme
  will become slightly complicated after such an replacement for one has to
  teleport $q$ary instead of binary quantum registers in
  step~\ref{Alg:Secret_Dist}. \label{Alg:EPR_Set}
 \item Let $x_i$ be the private classical input of player $i$, then he/she
  prepares $s = \mbox{O} (\log \frac{1}{e})$ copies of the state $|x_i\rangle$.
  He/She also prepares a number of preset quantum registers $|0\rangle$ that
  will be used later on in the reversible quantum computation. Player $i$ first
  encodes each of his/her prepared quantum registers using the QECC $C_1$.
  Then, player $i$ further encodes the $j$th quantum register in each of
  his/her encoded state using $C_2$ and teleports the resultant quantum
  registers to player $j$ using their previously shared encoded state $|\Phi
  \rangle$ in step~\ref{Alg:EPR_Set} for all $j\neq i$. He/She also encodes
  each of the $i$th quantum register by $C_2$ and keeps those quantum registers
  himself/herself. All players keep their received quantum registers private as
  well. And in what follows, I use the subscript ``${\rm L}$'' below the state
  ket to denote a state that is encoded and distributed among the $n$ players
  using this procedure. In addition, the players also prepare a number of
  preset quantum registers $|0\rangle$, encode it first by $\tilde{C}_1$ and
  then by $C_2$. The players then distribute these encoded preset registers
  among themselves in a similar way as in sharing their private inputs. And I
  use the subscript ``${\rm \tilde{L}}$'' below the state ket to denote such an
  encoded and distributed state. States $|0\rangle_{\rm L}$ and
  $|0\rangle_{\rm \tilde{L}}$ shall be used as preset registers during the
  reversible computation in step~\ref{Alg:Secret_Comp}. \label{Alg:Secret_Dist}
 \item In order to make sure that everyone follows step~\ref{Alg:Secret_Dist}
  honestly, a player $j$ (the verifier) may challenge a randomly chosen player
  $i$ (the prover) using the fault-tolerant random parity check method similar
  to that used in Ref.~\cite{qkd}.
  \par
  More precisely, player $j$ publicly announces a sequence $\{ c_k \}_{k=1}^s$
  of integers in ${\mathbb F}_q$ such that $\sum_{k=1}^s c_k = 0$. Then, every
  player is required to help player $j$ to compute the random parity
  $\sum_{k=1}^s c_k x_{ik}$ by distributed fault-tolerant quantum computation
  (FTQC), where $x_{ik}$ denotes the state of the $k$th copy of the private
  input of player $i$. Clearly, the choice of QECCs $C_1$ and $C_2$ enable us
  to perform the above quantum computation in a fault-tolerant way
  {\em without} any measurement and ancilla \cite{random}. Besides, the method
  of distributing the private input state in step~\ref{Alg:Secret_Dist} allows
  the players to perform the above FTQC in a distributed manner {\em without}
  any communications between them.
 \par
  To verify if the result computed (which I call it the random parity) is equal
  to zero, all players measure and publicly announce their measurement outcome
  along their commonly agreed computational basis on their corresponding $C_2$
  encoded quantum registers that encode the random parity. Because $C_1$ is a
  $[[n,1,d]]_q$ random polynomial QECC, the measurement results of the players
  correspond to the classical $[n,d,n\!-\!d\!+\!1]_q$ Reed-Solomon encoding of
  the random parity. Naturally, they continue only if the random parity
  inferred from this classical Reed-Solomon encoding is zero. This verification
  process has to repeat $\mbox{O} (\log \frac{1}{e})$ times for each proving
  player $i$ so as to guarantee security.
  \par
  In addition, all players use a similar distributed fault-tolerant random
  parity checking technique to verify the purity of the distributed encoded
  preset quantum registers $|0\rangle_{\rm L}$ and $|0\rangle_{\rm \tilde{L}}$
  among themselves. They proceed to step~\ref{Alg:Secret_Comp} only when all
  the measurement results are consistent with the assumption that there is no
  cheater or eavesdropper around. Thus, in order to establish the required
  security, $\mbox{O} (\log \frac{1}{e})$ private input states prepared and
  distributed in step~\ref{Alg:Secret_Dist} are wasted. (An alternative way to
  perform the random parity check measurement is to ask the players to teleport
  their shares of the encoded random parity quantum registers to the verifier.
  Then, the verifier makes the appropriate measurement and publicly announces
  the outcome.) \label{Alg:Verify}
 \item To compute the commonly agreed classical function $z=f(x_1,x_2,\ldots ,
  x_n)$, the $n$ players perform distributed FTQC on their received quantum
  particles. The players keep every quantum state except the final result
  private.
  \par
  To be precise, they first decompose the classical function $f$ into a
  commonly agreed composition of elementary operators. Each elementary operator
  is in the form of (i)~register-wise addition $|x\rangle \mapsto |x+a\rangle$,
  (ii)~register-wise multiplication $|x\rangle \mapsto |ax\rangle$,
  (iii)~generalized C-NOT $|x,y\rangle \mapsto |x,x+y\rangle$ and
  (iv)~generalized Toffoli gate $|x,y,z\rangle \mapsto |x,y,z+xy\rangle$, for
  some fixed $a\neq 0$ \cite{Barenco}.
  \par
  At this point, each player should have $r = \mbox{O} (\log \frac{1}{e}) < s$
  remaining quantum registers distributed among themselves. Moreover, all the
  remaining distributed quantum states of an honest player, upon quantum error
  correction, should be identical. Clearly, the choice of the random polynomial
  QECCs $C_1$ and $C_2$ together with the private secure distribution method in
  step~\ref{Alg:Secret_Dist} allow the players to perform the first three types
  of elementary operators without any measurement or communication between the
  players \cite{random}. Thus, they can perform the fault-tolerant operation on
  the $r$ remaining distributed quantum registers one by one. And in this way,
  they end up with having $r$ identical resultant states if they are honest.
  \par
  To perform the fourth type of elementary operator, namely, a generalized
  Toffoli gate on the $r$ remaining distributed encoded states, they do the
  following. First, the players collectively synthesize the distributed state
  $\sum_{a,b=0}^{q-1} |a,b,ab\rangle_{\rm L} / q^{3/2}$ among themselves using
  their verified distributed states $|0\rangle_{\rm \tilde{L}}$ by a procedure
  based on that in Ref.~\cite{Preskill} as follows:
  \begin{mathletters}
   \begin{eqnarray}
    & & |0,0,0,0\rangle_{\rm \tilde{L}} \nonumber \\
    & \longmapsto & \frac{1}{q^2} \sum_{a,b,c,k=0}^{q-1}
     |a,b,c,k\rangle_{\rm L} \label{E:Syn_Start} \\
    & \longmapsto & \frac{1}{q^2} \!\sum_{a,b,c,k=0}^{q-1} \!\omega_q^{-k c}
     |a,b,c,k\rangle_{\rm L} \label{E:Syn_CNOT} \\
    & \longmapsto & \frac{1}{q^2} \!\sum_{a,b,c,k=0}^{q-1} \!\omega_q^{k
     (ab-c)} |a,b,c,k\rangle_{\rm L} \label{E:Syn_CCPhase} \\
    & \longmapsto & \frac{1}{q^{5/2}} \!\!\sum_{a,b,c,k,x=0}^{q-1}
     \!\!\omega_q^{k (ab-c+x)} |a,b,c\rangle_{\rm L} \otimes
     |x\rangle_{\rm \tilde{L}} ~, \label{E:Syn_FFT}
   \end{eqnarray}
  \end{mathletters}
  where $\omega_q$ is a primitive $q$th root of unity.
 \par
  To arrive at Eq.~(\ref{E:Syn_Start}) in a fault-tolerant manner, each player
  $i$ simply has to perform the following local Fourier transformation
  $|a\rangle \longmapsto \sum_{b=0}^{q-1} \omega_q^{m_i a b} |b\rangle /
  \sqrt{q}$ on his/her corresponding quantum registers, where $m_i\in
  {\mathbb F}_q$ is a unique solution for the system of equations $\sum_{i=1}^n
  m_i = 1$ and $\sum_{i=1}^n m_i y_i = \sum_{i=1}^n m_i y_i^2 = \cdots =
  \sum_{i=1}^n m_i y_i^{n-1} = 0$. I denote this fault-tolerant transformation
  by ${\mathfrak F}$. In fact, Appendix~\ref{S:Transform_Note} shows that
  ${\mathfrak F} |0\rangle_{\rm L} = \sum_{k=0}^{q-1}
  |k\rangle_{\rm \tilde{L}}$ and ${\mathfrak F} |0\rangle_{\rm \tilde{L}} =
  \sum_{k=0}^{q-1} |k\rangle_{\rm L}$. And then, Aharonov and Ben-Or tell us
  how to arrive at Eqs.~(\ref{E:Syn_CNOT}) by fault-tolerant
  controlled-phase-shift gate {\em without} any communication between the
  players \cite{random}. More precisely, each player $i$ applies $|a,b\rangle
  \longmapsto \omega_q^{p_i ab} |a,b\rangle$ to their share of the third and
  fourth quantum registers where $p_i \in {\mathbb F}_q$ satisfies
  $\sum_{i=1}^n p_i = -1$ and $\sum_{i=1}^n p_i y_i = \sum_{i=1}^n p_i y_i^2 =
  \cdots = \sum_{i=1}^n p_i y_i^{2d} = 0$. Subsequently, arriving at
  Eq.~(\ref{E:Syn_CCPhase}) from Eq.~(\ref{E:Syn_CNOT}) requires the
  fault-tolerant controlled-controlled-phase-shift gate $|a,b,c\rangle_{\rm L}
  \mapsto \omega_q^{abc} |a,b,c\rangle_{\rm L}$. And for the random polynomial
  code $C_1$ with $3d\leq n\!+\!2$, this operation is achieved when each player
  $i$ applies the controlled-controlled-phase-shift gate $|a,b,c\rangle \mapsto
  \omega_q^{r_i abc} |a,b,c\rangle$ to his/her corresponding share of the
  encoded first, second and third quantum registers, where $r_i \in
  {\mathbb F}_q$ is the solution (not necessarily unique unless $3d\!+\!1 = n$)
  of the system of equations $\sum_{i=1}^n r_i = 1$, and $\sum_{i=1}^n r_i y_i
  = \sum_{i=1}^n r_i y_i^2 = \cdots = \sum_{i=1}^n r_i y_i^{3d} = 0$. Finally,
  to arrive at Eq.~(\ref{E:Syn_FFT}) from Eq.~(\ref{E:Syn_CCPhase}) in a
  fault-tolerant way, the players simply apply the same local Fourier transform
  ${\mathfrak F}$ that creates Eq.~(\ref{E:Syn_Start}) to their share of the
  fourth quantum register. (Again, the proof can be found in
  Appendix~\ref{S:Transform_Note}.) In summary, the players can evolve their
  share of quantum states to Eq.~(\ref{E:Syn_FFT}) in a fault-tolerant manner
  {\em without} any measurement, communications or the use of ancillary
  particles.
 \par
  After the players have evolved their quantum particles to the distributed
  state in Eq.~(\ref{E:Syn_FFT}), they measure their share of the fourth
  encoded quantum register along the commonly agreed computational basis and
  then publicly announce their measurement results. In this way, they end up
  having a classical $[n,n\!-\!d\!+\!1,d]_q$ Reed-Solomon code and after error
  correction, they can infer the measurement outcome of the fourth encoded
  quantum register along the commonly agreed computational basis. Suppose the
  inferred measurement result is $\lambda$, then the state ket of the remaining
  three distributed encoded quantum registers becomes $\sum_{a,b,c,k=0}^{q-1}
  \omega_q^{k (ab-c+\lambda)} |a,b,c\rangle_{\rm L} / q^2 = \sum_{a,b=0}^{q-1}
  |a,b,ab+\lambda\rangle_{\rm L} / q$. So, by applying a fault-tolerant
  generalized C-NOT gate depending on the measurement result $\lambda$, they
  eventually synthesize the state $\sum_{a,b=0}^{q-1} |a,b,ab\rangle_{\rm L} /
  q$ collectively.
  \par
  At this point, using their newly synthesized distributed encoded state
  $\sum_{a,b=0}^{q-1} |a,b,ab\rangle_{\rm L} / q$ as ancilla, the $n$ players
  implement the generalized Toffoli gate in a fault-tolerant manner using a
  variation of the Gottesman's method in Ref.~\cite{Gottesman2}. (See also
  Ref.~\cite{Preskill} for details.) More precisely, they perform the following
  transformation using a number of fault-tolerant generalized C-NOT gates and
  a fault-tolerant ${\mathfrak F}$ gate
  \begin{eqnarray}
   & & \frac{1}{q^2} \sum_{a,b,c=0}^{q-1} |x,y,z,a,b,ab\rangle_{\rm L}
    \nonumber \\
   & \longmapsto & \frac{1}{q^{3/2}} \sum_{a,b,c=0}^{q-1} \omega_q^{zc} |x-a,
   y-b\rangle_{\rm L} \otimes |c\rangle_{\rm \tilde{L}} \nonumber \\
   & & ~~~\otimes |a,b,z+ab\rangle_{\rm L} ~. \label{E:Toffoli}
  \end{eqnarray}
  Now, the $n$ players measure their shares of the first three encoded
  registers along the commonly agreed computational basis. Regarding as
  classical Reed-Solomon codes, their publicly announced measurement outcomes
  can then be used to infer the (quantum) measurement results of the first
  three registers along the commonly agreed computational basis. Suppose the
  inferred measurement results of the first three registers are $\lambda_1$,
  $\lambda_2$ and $\lambda_3$, respectively. Then, by adding $\lambda_1$ to the
  fourth register, $\lambda_2$ to the fifth register, and $\lambda_1
  y\!+\!\lambda_2 x\!-\!\lambda_1 \lambda_2$ to the sixth register, they get
  the state $\omega_q^{\lambda_3 z} |x,y,z+xy\rangle_{\rm L}$. Finally, they
  obtain the state $|x,y,z+xy\rangle_{\rm L}$, which is the result of a
  generalized Toffoli operation, by applying a suitable phase-shift gate in the
  sixth register and then followed by another controlled-controlled-phase-shift
  operator to the first and second registers. (As I have discussed previously,
  players may perform these operations without any communication because of the
  choice of the QECC $C_1$ and $C_2$ together with the fact that $\lambda_1$,
  $\lambda_2$ and $\lambda_3$ are classical data.)
  \par
  To ensure accuracy, they perform the above process $r$ times to the $r$
  supposedly identical signal states. In this way, they end up with
  implementing $r$ identical generalized Toffoli operators if all players
  are honest. (At this point, readers may wonder why I do not check the purity
  of ancillary state $\sum_{a,b=0}^{q-1} |a,b,ab\rangle_{\rm L} / q$ directly.
  The reason is that random parity checking does not work for this ancillary
  state because the state of the untested particles will be altered by the test
  itself. Readers may also ask why I do not apply the fault-tolerant Fourier
  transformation gate to obtain $\sum_{k=0}^{q-1} |k\rangle_{\rm L}$ from
  $|0\rangle_{\rm L}$. The reason is that all known fault-tolerant Fourier
  transformation gate for the $[[n,1,d]]_q$ QECC $C_1$ with $3d\leq n\!+\!2$ to
  date requires collective measurements on the encoded quantum registers and
  hence is liable to error in the presence of cheaters.) (An alternative method
  to perform the required measurement is to assign once and for all a randomly
  chosen player for each of the $r = \mbox{O} (\log \frac{1}{e})$ supposedly
  identical signal states. Whenever it comes to a measurement, players teleport
  their states to be measured to the corresponding assigned player who then
  makes the necessary measurement and publicly announces the measurement
  outcome.) \label{Alg:Secret_Comp}
 \item In order to make sure that the players indeed follow the distributed
  FTQC in step~\ref{Alg:Secret_Comp} honestly, they carry out the random parity
  verification test $\mbox{O} (\log\frac{1}{e})$ times to their final state
  using the same method as described in step~\ref{Alg:Verify}. Finally, to
  obtain the value of $z=f(x_1,x_2,\ldots ,x_n)$, the $n$ players separately
  measure their share of quantum registers that encodes the value of $z$ along
  the commonly agreed computational basis, and then publicly announce their
  measurement outcomes. Then, they infer the value of $z$ using standard
  classical Reed-Solomon code error correction. \label{Alg:Measure}
\end{enumerate}
\section{The Security Of The Quantum Scheme}
\label{S:Security}
 Now, I claim that the above scheme correctly computes the classical function
 $z=f(x_1,x_2,\ldots ,x_n)$ with a probability $1\!-\!\ell e$ for some fixed
 constant $\ell \geq 1$, provided that no more than $\delta$ players cheat.
 Besides, those $\delta \equiv \left\lfloor \frac{d-1}{2} \right\rfloor$
 cheaters know nothing about the private inputs of every honest player and they
 cannot alter the final outcome $z$. These claims are true even if all cheaters
 cooperate and have unlimited computational power.
\par
 To prove the above claims, one observes that there are four possible ways for
 the above scheme to go wrong, namely, the presence of noises, bad instruments,
 eavesdroppers and cheating players. Remember that a cheater may deliberately
 announce wrong measurement results and thereby misleading others. Besides, one
 has to make the most pessimistic assumption that all cheaters and
 eavesdroppers cooperate and control everything except the instruments in the
 laboratories of the honest players. The cheaters may even have unlimited
 computational power. Using the argument in Ref.~\cite{qkd}, I first show that
 we can safely neglect the effect of noises and bad instruments. Since all
 steps in the above scheme are performed in a fault-tolerant manner, the theory
 of FTQC tells us that with probability $1\!-\!e$ we may regard the effect of
 noise and bad instruments simply affect the error syndromes but not the
 quantum information encoded in the states \cite{random,ftqc,Preskill}.
 Besides, the theory of QECC tells us that learning error syndromes give no
 information about the quantum information encoded in the state \cite{qecc,KL}.
 Consequently, by restricting myself to the evolution of quantum information
 contained in the encoded quantum registers, I may analyze the behavior of the
 above scheme in a noiseless environment from now on.
\par
 Then, it remains for me to show that no more than $\delta$ cheaters can obtain
 partial information on the private inputs of some honest players. Besides,
 these cheaters cannot alter the output of the classical function $f$. In order
 to do so, one has to understand the function of each step in the scheme first.
 Steps~\ref{Alg:Relay_Selection} and~\ref{Alg:EPR_Set} are direct
 generalization of the entanglement-based quantum key distribution protocol
 proposed by Lo and Chau in Ref.~\cite{qkd}. The aim of these two steps is to
 share almost perfect encoded state $|\Phi\rangle$ between any two pairs of
 players so that they can teleport quantum states in a fault-tolerant manner
 from one to another at a later time in step~\ref{Alg:Secret_Dist}.
 Step~\ref{Alg:Verify} make sure that every player follows
 step~\ref{Alg:Secret_Dist} to distribute his/her private input as well as the
 preset quantum registers using the QECCs $C_1$ and $\tilde{C}_1$. The actual
 computation is carried out in step~\ref{Alg:Secret_Comp}. And finally, they
 verify and measure their computational result in step~\ref{Alg:Measure}.
\subsection{Private Inputs Of An Honest Player Is Secure Up To
 Step~\protect\ref{Alg:Verify} Of The Quantum Scheme}
\label{SS:Step-5}
 I have two cases to consider in order to show that the $\delta \equiv
 \left\lfloor \frac{d-1}{2} \right\rfloor$ cheaters obtain no information on
 the private inputs of the honest players up to the random parity verification
 in step~\ref{Alg:Verify} of the quantum scheme. The first case is when the
 proving player $i$ in step~\ref{Alg:Verify} is honest. In this case, the
 encoded state $|\Phi\rangle$ sharing scheme in step~\ref{Alg:EPR_Set} between
 the proving player $i$ and all other honest players is a straight-forward
 generalization of the quantum key distribution protocol of Lo and Chau in
 Ref.~\cite{qkd}. More importantly, as stated in Appendix~\ref{S:Hashing_Note},
 the random parity test in step~\ref{Alg:Verify} maps the basis ${\mathcal B} =
 \{ \sum_{k=0}^{q-1} \omega_q^{k b} |k,k+a\rangle / \sqrt{q} \}_{a,b\in
 {\mathbb F}_q}$ to basis ${\mathcal B}$ up to a global phase. Therefore, the
 proof of Lo and Chau in Ref.~\cite{qkd} applies. In particular, they have
 already proved that the fidelity of every encoded state $|\Phi\rangle$ shared
 between any two honest players is at least $1\!-\!e$ even in the presence of
 eavesdroppers and cheaters \cite{qkd}. Then in steps~\ref{Alg:Secret_Dist}
 and~\ref{Alg:Verify}, eavesdroppers and cheaters can only access to the public
 classical communications between the honest players. Fortunately, these
 classical messages contain no information about the teleported quantum state
 \cite{teleport}. Hence, no one apart from the sender and the receiver knows
 the teleported state. Thus, these $\delta$ cheaters have access to at most
 their share of $\delta$ quantum registers of the distributed encoded state
 $|x_i\rangle_{\rm L}$. Since the $C_1$ is a $[[n,1,d]]_q$ QECC, the knowledge
 of the $\delta$ quantum registers in the hands of the cheaters contains no
 information on the private input $x_i$ at all.
\par
 The second case is that the proving player $i$ is dishonest. Clearly, the job
 of the dishonest player $i$ is to somehow mislead the other players into
 believing that he/she is honest. More precisely, player $i$ tries to devise a
 method (possibly with the help of the other $\delta\!-\!1$ cheaters in the
 system) so as to pass the verification test in step~\ref{Alg:Verify} with a
 probability greater than $1\!-\!\ell e$ for some fixed positive constant
 $\ell$. Note that measuring every quantum register of an arbitrary quantum
 codeword of the $[[n,1,d]]_q$ random polynomial QECC $C_1$ along the commonly
 agreed computational basis gives a classical $[n,d,n\!-\!d\!+\!1]_q$
 Reed-Solomon codeword. Besides, if the $C_1$ encoded quantum state
 $|\Psi\rangle$ contains $\delta$ erroneous quantum registers, then after
 measuring along the computational basis, we end up getting a classical
 Reed-Solomon codeword with at most $\delta$ erroneous registers. Since $\delta
 < n/4$ \cite{KL,Chau}, therefore if an error can be handled by the QECC $C_1$,
 the corresponding error after measurement can be handled by the corresponding
 classical Reed-Solomon code. Moreover, the coarse-grained measurement, that
 is, process of measuring each quantum register along the computational basis
 together with the inference of quantum state from the Reed-Solomon code, can
 be regarded as a projective measurement along the $C_1$ encoded computational
 basis on quantum state. And now in the verification step~\ref{Alg:Verify}, all
 the $n\!-\!\delta$ honest players indeed measure the quantum states along the
 commonly agreed computational basis. Besides, the random parity check does not
 alter the state of the un-measured quantum particles. Therefore, the
 coarse-grained measurements performed by the honest players commute with each
 other; and hence each coarse-grained measurement result will in no way change
 the outcome of all subsequent measurements \cite{qkd}. Thus, theoretically,
 the honest players may push their coarse-grained measurement forward to the
 time when the quantum states are just prepared. Consequently, the probability
 that cheating player $i$ passes the quantum verification test in
 step~\ref{Alg:Verify} cannot exceed the probability of passing a classical
 random parity verification test in which player $i$ is only allowed to prepare
 only a classical mixture of states \cite{qkd}. Clearly, the probability that
 player $i$ cheats and yet he/she passes the classical verification test is no
 greater than $1/q^r$ where $r$ is the number of independent rounds of tests
 performed. Consequently, by repeating the quantum random parity test $\log_q
 \frac{1}{e}$ times, the probability that player $i$ cheats and yet he/she
 passes the quantum verification test in step~\ref{Alg:Verify} is at most $e$.
 And once the quantum verification test is passed, the fidelity of the
 remaining untested quantum states as being a valid input $|x_i\rangle$ is
 equal to $1\!-\!\ell e$ for some constant $\ell$ independent of $n$ and $e$.
 Thus, the entropy of each of the untested quantum states is equal to $\log
 q\!+\!\ell e$. Hence, the cheaters have exponentially small amount information
 on the private inputs of every honest player \cite{qkd}. And using a similar
 argument, I know that the fidelity of the distributed preset quantum registers
 $|0\rangle_{\rm L}$ and $|0\rangle_{\rm \tilde{L}}$ is also equal to
 $1\!-\!\ell e$.
\par
 Therefore, I conclude that if there are at most $\delta$ cheaters around and
 that they choose to perform measurements individually, then the probability
 that these cheaters can obtain partial information on the private inputs of
 the honest players is bounded from above by $\ell e$ for some fixed constant
 $\ell > 0$ up to step~\ref{Alg:Verify} of the quantum scheme.
\par
 In the event that the players choose to teleport their random parity state to
 the verifier who then make the necessary measurement, the proof of security up
 to step~\ref{Alg:Verify} is similar. Note that if the verifier is honest, then
 the above proof applies. On the other hand, if the verifier cheats, two
 possible things may happen. First, the verifier may wrongly announce an
 inconsistent result. But leads to an immediate abortion of the scheme. Hence,
 he/she cannot obtain any extra information on the private input of an honest
 player. Second, the verifier may turn a blind eye to a measurement result
 that is inconsistent with the no cheater/eavesdropper assumption. Since
 $\delta / n < 1/6$, a non-zero fraction of the verifiers are honest. So, after
 $\mbox{O} (\log \frac{1}{e})$ rounds of random parity tests, the probability
 that the private input of an honest players leaks out is less than $\ell e$
 for some fixed constant $\ell > 0$ up to step~\ref{Alg:Verify} of the quantum
 scheme.
\par
 Thus, I conclude that if there are at most $\delta$ cheaters around and that
 the players choose to teleport the particles encoding the random parities to
 the verifiers before making measurement, then the probability that cheaters
 obtain partial information on the private input of an honest player is less
 than $\ell e$ for some fixed constant $\ell > 0$.
\subsection{Cheater Cannot Alter The Computation Result}
\label{SS:Computation_Security}
 Now, I proceed to show that these $\delta$ cheaters cannot alter the outcome
 of the function evaluation $f$ with a probability greater than $e$ in
 steps~\ref{Alg:Secret_Comp} and~\ref{Alg:Measure} of the quantum scheme. Since
 one may regard any illegal quantum manipulation by the $\delta$ cheaters as
 decoherence acting on up to $\delta$ quantum registers in the QECC $C_1$, the
 theory of FTQC implies that any {\em quantum} manipulation by these cheaters
 cannot alter the final outcome of the function $f$. Nevertheless, the theory
 of FTQC assumes that all measurements of the encoded quantum state and
 manipulation of classical data are error free. So, it remains for me to show
 that measurement and classical data manipulation by cheaters also cannot alter
 the outcome of the function $f$.
\par
 Because of the choice of $C_1$ and $C_2$, there are two possible operations in
 the scheme that requires measurement or classical message communication,
 namely, the verification test and the generalized Toffoli gate. As I have
 discussed previously, incorrect measurement or classical message broadcasting
 in a verification test results in the immediate abortion of the scheme. Hence,
 it cannot alter the final output of the function $f$. So, it remains for me to
 consider to case of a generalized Toffoli gate. Recall that the generalized
 Toffoli gate is collectively synthesized by the $n$ players from the verified
 distributed encoded state $|0\rangle_{\rm \tilde{L}}$ in
 step~\ref{Alg:Secret_Comp}. Fortunately, if the players choose to perform
 their measurements individually, then all measurement results in
 step~\ref{Alg:Secret_Comp} are in either the $[n,d,n\!-\!d\!+\!1]_q$ or the
 $[n,n\!-\!d,d]_q$ Reed-Solomon code forms. Hence, the $\delta$ cheaters cannot
 alter the measurement outcome and hence the value of $z$.
\par
 On the other hand, if they choose to teleport their states to their
 corresponding randomly assigned player, then in order to pass the final random
 parity test in step~\ref{Alg:Measure} with a probability greater than $e$, the
 cheaters must arrange the state of the final outcome $z=f(x_1,x_2,\ldots
 ,x_n)$ for each of the $r = \mbox{O} (\log \frac{1}{e})$ copies of quantum
 particles to be almost identical. This is possible only when all the $r$
 randomly assigned players who are responsible for measurement cheat. Since the
 probability that all randomly assigned players cheat is equal to $\left(
 \frac{\delta}{n} \right)^r = \mbox{O} (e)$. Consequently, the probability that
 the $\delta$ cheaters can alter the final value of $z$ without being detected
 is equal to $\ell e$ for some fixed positive constant $\ell$.
\subsection{Cheater Cannot Obtain Partial Information During Distributed
 Computing Of The Function $f$}
\label{SS:Inform_Distributed}
 Although cheaters cannot alter the final outcome of the computation with a
 probability greater than $\ell e$ for some fixed positive constant $\ell$,
 readers may ask if these cheaters can obtain partial information on the
 private input of an honest player in steps~\ref{Alg:Secret_Comp}
 and~\ref{Alg:Measure}. Now, I show that this is not possible. Using the same
 argument as in Subsection~\ref{SS:Computation_Security} together with the
 choice of $[[n,1,d]]_q$ codes $C_1$ and $C_2$, the only possible place for
 information leakage is the measurement performed by the players during
 the implementation of a generalized Toffoli gate. And as I have discussed in
 Subsection~\ref{SS:Computation_Security}, if the players choose to measure
 individually, then the $\delta$ cheaters cannot alter the joint measurement
 result that is required during the collective and distributive synthesis of
 the ancillary state $\sum_{a,b=0}^{q-1} |a,b,ab\rangle_{\rm L} / q$ as well as
 during the implementation of the generalized Toffoli gate. Moreover, theory of
 QECC tells us that the value of these measurements contains no information on
 the distributed encoded state $|x,y,z\rangle_{\rm L}$. Recall that the
 $\delta$ cheaters have access only to their shares of the entangled quantum
 state together with the classical information on the measurement results on
 the fault-tolerant generalized Toffoli gate. Since $C_1$ is a $[[n,1,d]]_q$
 QECC, these information alone is not enough for the cheaters to obtain any
 information on $|x,y,z\rangle_{\rm L}$ and hence the private inputs of an
 honest player.
\par
 On the other hand, if the players choose to teleport their corresponding
 states to the randomly assigned players before making measurements, then we
 cannot control the action of a cheating assigned player. Nonetheless, by
 looking into the synthesis scheme of the ancillary state $\sum_{a,b=0}^{q-1}
 |a,b,ab\rangle_{\rm L}$ used in step~\ref{Alg:Secret_Comp}, the cheating
 assigned player can only alter the third encoded quantum register of this
 ancillary state. In other words, the cheating assigned player can only, after
 error correction, alter the state of the last quantum register in
 Eq.~(\ref{E:Toffoli}). So right after all players teleported their
 corresponding quantum registers to the cheating assigned player, the $\delta$
 cheaters control the first three encoded quantum registers together with the
 shares of distributed encoded fourth, fifth and sixth registers. Consequently,
 the reduced density matrix of the quantum registers controlled by the cheating
 assigned players is independent of $x$, $y$ and $z$. Hence, it is impossible
 for the $\delta$ cheaters to obtain partial information of the private input
 of an honest player.
\par
 In summary, using the results in
 Subsections~\ref{SS:Step-5}--\ref{SS:Inform_Distributed}, I conclude that the
 quantum secure multi-party computation scheme in Section~\ref{S:Scheme} is
 secure provided that no more than $\delta$ players cheat. Moreover, the
 security is unconditional for it does not rely on any computational
 assumption.
\par
 And in the alternative scheme that the players teleport their quantum states
 to some once and for all randomly chosen players and let these assigned
 players to make the measurement, the proof that the $\delta$ cheaters cannot
 alter the final outcome $z$ and that they cannot obtain extra information on
 the private input of an honest player is similar.
\section{The Complexity And Security Tradeoff Between The Quantum And Classical
 Schemes}
\label{S:Complexity}
 Clearly, the above quantum secure multi-party computation scheme requires
 $\mbox{O} (n)$ quantum channels, a public classical unjamable broadcasting
 channel, $\mbox{O} (n^2 k\log \frac{1}{e})$ bits of quantum and classical
 communications in order to distribute and compute the classical function $f$,
 where $k$ is the length of each private input. Distributed FTQC of
 register-wise addition, register-wise multiplication and generalized C-NOT
 gate do not require any communication. And distributed FTQC of a generalized
 Toffoli gate requires $\mbox{O} (nk\log\frac{1}{e})$ bits of classical
 messages broadcast, or equivalently, $\mbox{O} (n^2 k\log\frac{1}{e})$ bits of
 classical communications between the players if they choose to perform their
 measurement individually. Distributed FTQC of a generalized Toffoli gate
 requires $\mbox{O} (nk\log \frac{1}{e})$ bits of classical communications
 should they choose to teleport the states and measure them collectively by the
 randomly assigned players. Moreover, if classically non-distributed computing
 $f$ requires $T$ timesteps and $S$ space, then the distributed quantum
 computing scheme in step~\ref{Alg:Secret_Comp} above requires $\mbox{O} (n
 T^{1+\epsilon})$ timesteps and $\mbox{O} (n S\log T)$ space for any $\epsilon
 > 0$ \cite{reversible}. Hence, the amount of communication required to
 distributed FTQC of a classical function $f$ is bounded from above by
 $\mbox{O} ( n^2 k T^{1+\epsilon} \log \frac{1}{e})$ should they use the
 alternative teleportation plus measurement method. In contrast, the best
 classical secure multi-party computation scheme known to date requires
 $\mbox{O} (n^2)$ communication channels and $\mbox{O} (n^3 k T)$ bits of
 communications. Thus, the quantum secure multi-party computation scheme
 requires fewer channels and less computation or communications than the best
 known classical algorithm to date.
\par
 Nevertheless, the improvement of the quantum scheme over the classical one
 comes with a price tag. Recall that the maximum number of cheaters tolerated
 by this quantum scheme is related to the maximum possible distance $d$ of a
 QECC that maps one $q$ary quantum register to $n$ $q$ary quantum registers.
 Since I am using the $[[n,1,d]]_q$ QECC with $3d \leq n\!+\!2$, my scheme can
 tolerate only asymptotically up to strictly less than 1/6 cheaters. On the
 other hand, the best known classical scheme is unconditionally secure provided
 that strictly greater than one half of the players are honest. In other words,
 the quantum scheme reported here trades security for communication complexity.
\section{Full-Blown Quantum Code Is Required In The Quantum Scheme}
\label{S:Code}
 At this point, readers may question if a full-blown QECC is required in this
 quantum scheme because phase errors do not affect the final outcome $z$.
 Rather surprisingly, the answer is yes. In fact, I shall show that if $C$ is a
 linear map sending one quantum register to $n$ quantum registers, then any two
 of the three conditions below imply the third one:
\begin{enumerate}
 \item $C$ is a QECC correcting up to $\delta$ spin flip errors.
 \item $C$ is a QECC correcting up to $\delta$ phase shift errors.
 \item The partial trace over any $n\!-\!\delta$ registers gives no information
  on the initial unencoded wavefunction.
\end{enumerate}
\par\indent
 The theory of QECC implies that (1) and (2) $\Rightarrow$ (3). And now, I show
 that (1) and (3) $\Rightarrow$ (2). The remaining case that (2) and (3)
 $\Rightarrow$ (1) can be proven in a similar way. I divide the $n$ players
 into two groups. Groups A and B have $n\!-\!\delta$ and $\delta$ players,
 respectively. By Schmidt polar decomposition, the encoded normalized state
 $\sum_k \alpha_k |k\rangle_{\rm L}$ can be written as $\rho = \sum_{i,j,k,k'}
 \alpha_k \overline{\alpha}_{k'} \sqrt{\lambda_i (k) \lambda_j (k')} |a_i (k)
 \rangle\otimes |b_i (k)\rangle \langle a_j (k')|\otimes\langle b_j (k') |$,
 where $|a_i (k)\rangle$ and $|b_i (k)\rangle$ are eigenvectors of the reduced
 density matrices as seen by groups A and B, respectively. Hence, taking
 partial trace over group A, condition~(3) tells us that
\begin{equation}
 {\rm Tr}_A (\rho) = \sum_{i,j,k,k'} \alpha_k \overline{\alpha}_{k'} \langle
 a_j (k') | a_i (k) \rangle \,|b_i (k)\rangle\langle b_j (k') | \label{E:Indep}
\end{equation}
 is independent of $\alpha_k$. This is possible only if $|b_i (k)\rangle \equiv
 |b_i\rangle$ and $\sqrt{\lambda_i (k) \lambda_j (k')} \langle a_j (k') | a_i
 (k) \rangle$ are independent of $k$ for all $i,j$. Condition~(1) implies that
\begin{equation}
 \sum_{i,j} \sqrt{\lambda_i (k) \lambda_j (k')} \langle b_i | S^{\dag} | b_j
 \rangle \,\langle a_i (k) | S' | a_j (k') \rangle = \delta_{k,k'}
 \Lambda_{S,S'}, \label{E:kk'}
\end{equation}
 where $S$ and $S'$ are spin flip operators such that each acts on no more than
 $\delta$ quantum registers, and $\Lambda_{S,S'}$ is independent of $k$ and
 $k'$ \cite{qecc,KL}. Since $|b_i\rangle$ is independent of $k$,
 Eq.~(\ref{E:kk'}) holds if one replaces $S$ by a general quantum error
 operator $G$ which acts on no more than $\delta$ quantum registers. Since
 groups A and B are arbitrarily chosen, Eq.~(\ref{E:kk'}) is valid if one
 replaces $S'$ by $G$. Once again, since $|b_i\rangle$ is independent of $k$,
 I conclude that Eq.~(\ref{E:kk'}) is true even if one replaces the two spin
 flip operators $S$ and $S'$ by general quantum error operators $G$ and $G'$
 which act on no more than $\delta$ quantum registers. Consequently, $C$ is a
 QECC correcting up to $\delta$ errors \cite{qecc,KL}. In particular,
 condition~(2) is valid.
\section{Outlook}
\label{S:Outlook}
 In summary, I have reported and proved the security of a quantum secure
 multi-party scheme to compute classical functions. The scheme makes essential
 use of fault-tolerant quantum computation and a specially designed quantum
 error correcting code. While the quantum scheme tolerates only about one third
 the number of cheaters as the best known classical scheme to date, it requires
 asymptotically smaller amount of communication between the players.
\par
 This scheme also tells us that higher dimensional CSS-like quantum error
 correcting codes with fault-tolerant implementation have far-reaching
 applications outside the context of quantum mechanical computation. While
 quantum code is not the only possible way to protect quantum information
 during computation \cite{subspace}, cheating players may do all the nasty
 things that only full-blown quantum code can handle. Hence, quantum code is an
 essential ingredient in this secure multi-party computation scheme. Moreover,
 no binary $[[n,1,d]]_2$ CSS code with $d > n/7$ is known to date. Thus, higher
 dimensional quantum code \cite{nary} appears to be an essential ingredient in
 making my scheme to tolerate strictly less than one sixth cheating players.
 Since fault-tolerant computation of a general non-CSS-like code requires
 collective measurements \cite{Gottesman2}, it seems likely that $C_1$ should
 be a CSS-like code \cite{css}. Besides, by replacing the random polynomial
 codes $C_1$ and $C_2$ by the corresponding continuous quantum codes
 \cite{continuous} of the form $|a_0\rangle \longmapsto \int da_1 \, da_2
 \cdots \, da_{d-1} \bigotimes_{i=1}^n |a_0 + a_1 y_i + \cdots + a_{d-1}
 y_i^{d-1}\rangle$, my scheme also works for continuous quantum variables.
\par
 Rains showed that no binary $[[n,1,2\delta\!+\!1]]_2$ quantum code exists for
 $\delta > n\!+\!1$ \cite{Shadow} and a simple modification of the proofs of
 the optimality of the five quantum register code in Refs.~\cite{KL}
 and~\cite{Chau} shows that $[[n,1,d]]_q$ codes must satisfy $d/n < 1/4$. Thus,
 it may be possible to design a QECC based secure multi-parity computation
 scheme that tolerates up to a quarter cheaters. It is instructive to find such
 scheme out, if any.
\par
 It is also natural to ask if it is possible to extend this scheme to perform
 multi-party computation of a {\em quantum function}. That is, given a commonly
 agreed unitary operator $U$ as well as $n$ private quantum states
 $|x_i\rangle$, is it possible to compute $U \otimes_i |x_i\rangle$? Clearly,
 such a scheme exists if all the players are honest. The players may simply
 modify the scheme in this Paper a little bit by dropping out all the
 verification tests that check the identity of the private inputs, final
 output, and the correct implementation of generalized Toffoli gates.
 Nevertheless, there is no obvious way to use the random parity test to check
 the validity of a general quantum state. Moreover, a player may cheat by using
 the delay measurement tactics as in the proof of the impossibility of quantum
 bit commitment \cite{Bit_Comm}. It is, therefore, of great interest to know if
 it is possible to achieve quantum multi-party computation of a quantum
 function in the presence of cheaters.
\appendix
\section{Procedure Of Teleporting A $q$ary State}
\label{S:Teleport_Note}
 The $q$ary state quantum teleportation process goes as follows: The sender and
 the receiver first share the state $|\Phi\rangle = \sum_{k=0}^{q-1} |kk\rangle
 / \sqrt{q}$ before the sender makes a joint measurement on the quantum state
 $|\Psi\rangle$ to be teleported and his/her share of the state $|\Phi\rangle$
 along the basis $\{ \sum_{k=0}^{q-1} \omega_q^{b k} |a,a+k\rangle / \sqrt{q}
 \}_{a,b\in {\mathbb F}_q} \}$ where $\omega_q$ is a primitive $q$th root of
 unity. Then, the sender informs the receiver the measurement result. If the
 measurement outcome is $\sum_{k=0}^{q-1} \omega_q^{b k} |a,a+k\rangle /
 \sqrt{q}$, then the receiver may reconstruct the quantum state $|\Psi\rangle$
 by applying the unitary transformation $|x\rangle \mapsto \omega_q^{b (x-a)}
 |x-a\rangle$ to his/her share of the original state $|\Phi\rangle$.
\section{Procedure Of The Random Parity/Hashing Test}
\label{S:Hashing_Note}
 Let us consider the basis ${\mathcal B} = \{ \sum_{k=0}^{q-1} \omega_q^{k b}
 |k,k+a\rangle / \sqrt{q} \}_{a,b\in {\mathbb F}_q}$. Clearly, one may
 transform from one basis state ket to another by local unitary operations
 alone. And I denote the set of all such transformations by $T$. Furthermore,
 the register-wise generalized C-NOT operation maps the basis states
 ${\mathcal B} \otimes {\mathcal B} \equiv \{ |A\rangle \otimes |B\rangle :
 |A\rangle, |B\rangle \in {\mathcal B} \}$ to ${\mathcal B} \otimes
 {\mathcal B}$ up to a global phase. Therefore, the random parity/hashing test
 goes as follows: the two parties cooperate and randomly apply a transform
 $f_i \in T$ for each share of their entangled quantum state they obtain in
 step~\ref{Alg:EPR_Set}. Then they apply the register-wise generalized C-NOT
 operations to a number of randomly selected pairs of their resultant entangled
 quantum states. Finally, they measure the outcome of their final target
 quantum register along the computational basis. They continue only if their
 measurement result is consistent with the hypothesis that their share of
 quantum particles are all in the state $|\Phi\rangle$. And if they continue,
 they apply suitable transformations $g_i \in T$ on their remaining shares of
 quantum states so as to bring them back to the state $|\Phi\rangle$. Clearly,
 this random parity checking procedure is a direct generalization of that used
 in Ref.~\cite{qkd}.
\section{The Action Of ${\mathfrak F}$}
\label{S:Transform_Note}
 Here I show that ${\mathfrak F} |0\rangle_{\rm L} = \sum_{k=0}^{q-1}
 |k\rangle_{\rm \tilde{L}}$. The proof of ${\mathfrak F}
 |0\rangle_{\rm \tilde{L}} = \sum_{k=0}^{q-1} |k\rangle_{\rm L}$ is similar.
 Recall that ${\mathfrak F}$ denotes the collective action of $|a,b\rangle
 \longmapsto \omega_q^{m_i ab} |a,b\rangle$ by the $i$th player on their share
 of the encoded quantum registers, where $m_i \in {\mathbb F}_q$ satisfies the
 system of equations $\sum_{i=1}^n m_i = 1$ and $\sum_{i=1}^n m_i y_i =
 \sum_{i=1}^n m_i y_i^2 = \cdots = \sum_{i=1}^n m_i y_i^{n-1} = 0$. Thus,
\begin{eqnarray}
 & & {\mathfrak F} |a_0\rangle_{\rm L} \nonumber \\
 & = & \!\!\!\!\!\sum_{a_1,a_2,\ldots ,a_{d-1},b_0,b_1,\ldots ,b_{n-1}=0}^{q-1}
  \!\!\!\!\!\omega_q^{\sum_{i=1}^n \sum_{j=0}^{d-1} \sum_{k=0}^{n-1} m_i a_j
  b_k y_i^{j+k}} \nonumber \\
 & & ~~\bigotimes_{i=1}^n |b_0 + b_1 y_i + \cdots + b_{n-1} y_i^{n-1} \rangle
  ~. \label{E:Sum_F}
\end{eqnarray}
\par\indent
 Summing over $a_1$ in Eq.~(\ref{E:Sum_F}) gives $b_{n-1} = 0$. And then
 summing over $a_2$ gives $b_{n-2} = 0$. And inductively, I conclude that
 Eq.~(\ref{E:Sum_F}) becomes $\sum_{b_0,b_1,\ldots ,b_{n-d}} \omega_q^{a_0 b_0}
 \bigotimes_{i=1}^n |b_0 + b_1 y_i + \cdots + b_{n-d} y_i^{n-d} \rangle$.
 Hence, by putting $a_0 = 0$, I obtain ${\mathfrak F} |0\rangle_{\rm L} =
 \sum_{k=0}^{q-1} |k\rangle_{\rm \tilde{L}}$, which is our required result.
\acknowledgments
 I would like to thank Debbie Leung for her valuable discussions and H.-K.~Lo
 for his useful suggestions to improve my presentation. Moreover, very useful
 discussions with C.~Cr\'{e}peau on the relation between random polynomial
 quantum code and classical Reed-Solomon code during the Quantum Computation
 Workshop in Isaac Newton Institute, Cambridge is gratefully acknowledged. This
 work is supported by the Hong Kong Government RGC grants HKU~7095/97P and
 HKU~7143/99P.

\end{multicols}
\end{document}